\documentclass{elsart}
\usepackage{epsfig}

\begin{document}
\begin{frontmatter}

\title{Study of the photon flux from the night sky at La Palma and Namibia,
in the wavelength region relevant for imaging atmospheric Cherenkov telescopes}

\author{S. Preu\ss,}
\author{G. Hermann,}
\author{W. Hofmann,}
\author{A. Kohnle}
\address{Max-Planck-Institut f\"ur Kernphysik,
D 69029 Heidelberg, P.O. Box 103980}

\begin{abstract}
The level of the night sky background light at La Palma and Namibia was
determined, with emphasis on the wavelength region 
and solid angle coverage relevant for
the operation of imaging atmospheric Cherenkov telescopes. The dependence
of the night sky background light both on celestial coordinates ($alt, az$)
and on galactic coordinates ($b, l$) was measured, with an angular resolution
of about $1^\circ$. Average light levels near the zenith are 
similar in both locations
-- $2.2 \cdot 10^{12}$ to $2.6 \cdot 10^{12}$ photons per sr~s~m$^2$ for
300~nm~$< \lambda <$~650~nm. With increasing zenith angle the level of
background light increases at La Palma, whereas a constant level is 
measured in Namibia.
Near the center of the Milky Way, background light levels are increased
by a factor up to 4 and more. 
Also the level
of light backscattered from the ground has been studied.
\end{abstract}


\end{frontmatter}

\section{Introduction}

Imaging atmospheric Cherenkov telescopes (IACTs) have emerged as the most
powerful instrument for gamma-ray astronomy in the TeV energy regime.
IACTs image the Cherenkov light emitted by an air shower 
onto a photomultiplier (PMT) camera. The orientation of the image in the camera
is related to the direction of the air shower, the intensity to the energy
of the primary, and the shape of image can be employed to separate 
electromagnetic showers induced by high-energy gamma-rays from those generated
by cosmic-ray nuclei interacting in the atmosphere.  

The typical yield of Cherenkov photons is roughly 100 photons~m$^{-2}$TeV$^{-1}$,
which arrive with a time dispersion of a few ns. Given this modest light yield,
the light of the night-sky background, with intensities of order 
$10^{12}$ photons~m$^{-2}$s$^{-1}$sr$^{-1}$, represents 
a significant limitation for the 
performance of IACTs. It defines the minimum Cherenkov light yield which 
can be reliably detected with the instrument, and hence its energy threshold.
Noise in the Cherenkov image due to background light from the night sky
results in additional uncertainties in the determination of image orientation,
image intensity, and image shape. Finally, the background light generates a 
continuous DC current in the photon detectors, which may limit their lifetime
and which in the case of photomultipliers may require the operation at modest 
gain, in order to limit the total charge accumulated on the last dynodes and
the anode during the lifetime of the experiment. Variations of the night
sky background light across the sky may also cause changes in the detection
efficiencies, in particular once tight cuts are applied to select gamma-ray
candidates, and may result in a systematic bias in ON-OFF measurements,
where the rate from a background region of the sky is used to estimate 
the rate of background events in the signal region; care has to be taken
to select signal and background regions with an identical sky brightness.

A number of different components contribute to the night sky 
background light and its variation with celestial and terrestrial
coordinates, here given roughly in the order of their importance:
\begin{itemize}
\item Night airglow, which results from photochemical processes of ions
in the upper atmosphere; its intensity increases with increasing zenith
angle.
\item The zodiacal light, which arises from the scattering of sun light by
interplanetary dust near the ecliptic, and is most prominent shortly
after sunset in the west, or before sunrise in the east.
\item Starlight, which is partially resolved into discrete stars.
The density of stars and star light varies by a large factor between
the Milky Way and other regions of the sky.
\item Diffuse galactic light, resulting from the scattering
of star light by interstellar dust, in particular in the
galactic plane.
\item Extragalactic light, from unresolved galaxies and light scattered
in intergalactic space represents only a very small contribution.
\item Aurora, which is, however, negligible at low latitudes.
\end{itemize}
A detailed discussion of these various light sources is given 
in \cite{ROA73}; a more recent exhaustive compilation of data 
as well as further references can be 
found in \cite{Leinert}. Specific results for La Palma --
one of the sites investigated -- are given in \cite{BEN98}.

In addition to these natural light sources, artificial light 
scattered in the atmosphere contributes both a continuum as well
as discrete lines (Na, Hg), see \cite{Garstang} for details.

Currently, a new stereoscopic IACT system - the High Energy Stereoscopic
System, or H.E.S.S. (\cite{AHA97,Ho1,Ho2}) - is under construction in the Khomas Highland of 
Namibia. H.E.S.S. consists initially of four telescopes, with mirror
areas of roughly 100 m$^2$ each, 15~m focal length, and cameras consisting
of 960 photomultiplier tubes covering a field of view of $5^\circ$ diameter. The four
telescopes provide multiple views of the same shower from different 
perspectives, significantly enhancing the ability to reconstruct the
orientation of the shower axis in space, to determine the shower energy
and to reject cosmic-ray background showers. 
To assist in the layout of the camera and its trigger system
and in the planning of
observations, knowledge of the level of night-sky background is
desirable.

While the night-sky background light has been extensively
investigated and documented (see, e.g., \cite{Leinert}), 
it is not straight-forward to use these results to estimate
the background seen by the photomultipliers of a Cherenkov
telescope; one reason is that typical measurements refer to the
dark sky, with no detectable stars in CCD images of large
astronomical telescopes. Cherenkov telescopes with their
relatively large image pixels will almost always integrate over a number
of faint stars. Also, while the visibility conditions
in the Gamsberg region of Namibia -- near the H.E.S.S. site --
have been explored extensively in earlier campaigns
\cite{Neckel,Eso}, little information is available 
about the night-sky background, and in particular about the 
local light pollution.

Therefore, an instrument was developed to reliably measure the
yield of night sky photons in the wavelength range relevant for
Cherenkov telescopes, and measurement campaigns were conducted at
La Palma -- the location of the HEGRA telescopes \cite{hegra1,hegra2} -- and in the 
Khomas Highland of Namibia. Results are reported in this paper.
The following sections introduce
the instrument and its calibration, discuss the results obtained in the two
campaigns, and conclude with the implications for the H.E.S.S.
telescopes.

A caveat to be mentioned is that the various components of the 
night-sky background light depend in differing ways on celestial
and terrestrial coordinates and on time -- starlight and diffuse
galactic light is a function of galactic coordinates; 
airglow is primarily a function of terrestrial coordinates,
and shows short-term as well as seasonal and solar-cycle
related time variations; the zodiacal
light depends on ecliptic coordinates; the contribution from
artificial light sources varies with terrestrial coordinates and
atmospheric conditions. Light is attenuated and scattered along
its path through the atmosphere, influenced by local and large-scale
atmospheric conditions. With only ground-based measurements at a given
location and during a short time interval, one cannot always 
identify these individual contributions; it is non-trivial
to interpret the results, and to judge to which extent they can be
generalized. We believe, however, that data were taken under rather typical
conditions, and would not expect the main conclusions to change as a 
result of a more detailed and longer-term study.

\section{The instrument and its calibration}

In the design of an instrument to monitor the brightness of the night sky,
two key choices concern the wavelength region covered, and the
solid angle over which the instrument averages the photon flux. 

The 
wavelength regime relevant for the detection of Cherenkov light is 
defined by the $1/\lambda^2$ distribution of the Cherenkov photons,
combined with the atmospheric transmission and the quantum efficiency
of the photodetectors.
At short wavelengths, atmospheric transmission introduces a cutoff
at about 300~nm. At long wavelengths, beyond about 500~nm, the quantum
efficiency of conventional photocathodes deteriorates rapidly, and the 
contributions from longer wavelengths to the photoelectron yield are
small. In order to directly match the sensitivity of the photon detectors
of the H.E.S.S. Cherenkov cameras \cite{kohnle_beaune}, a photomultiplier with a
bi-alkali photocathode
was used for the night-sky measurements. In order to test and verify the
wavelength dependence of the light yield, optional narrow-band filters
of 360~nm, 400~nm, 450~nm and 500~nm could be introduced in the light path.

Concerning the solid angle coverage, the relevant scales are set by 
the field of view of the Cherenkov cameras, with around $4^\circ$ to
$5^\circ$ diameter, and by the field of view of the individual image
elements - photomultiplier ``pixels'' with characteristic diameters
in the range of $0.1^\circ$ to $0.4^\circ$; the H.E.S.S. telescopes will
use a $0.16^\circ$ pixel size. The relevant level of the night-sky
background is given by the average photoelectron rate induced in these
pixels from dark regions of the sky; bright stars, which light up
individual pixels can be handled separately. In case of the H.E.S.S.
telescopes, stars between mag. 2 and 5, depending on the spectrum, will
cause doubling of the 
typical PMT DC current, and will be clearly visible in the camera~\footnote{We
note that since most Cherenkov cameras use AC-coupled readout 
electronics for the Cherenkov signals, the shift in the DC current
due to a star imaged onto a pixel will not bias the measurement 
of the Cherenkov signal, as long as the PMT remains in the region
of linear response; the starlight will, however, contribute increased
fluctuations, i.e., noise.}. For practical reasons, it was decided
to enlarge the acceptance of the instrument somewhat compared to 
individual pixels, and a field of view of about $0.8^\circ$ effective diameter
was chosen.

\begin{figure}
\begin{center}
\mbox{
\epsfxsize10.0cm
\epsffile{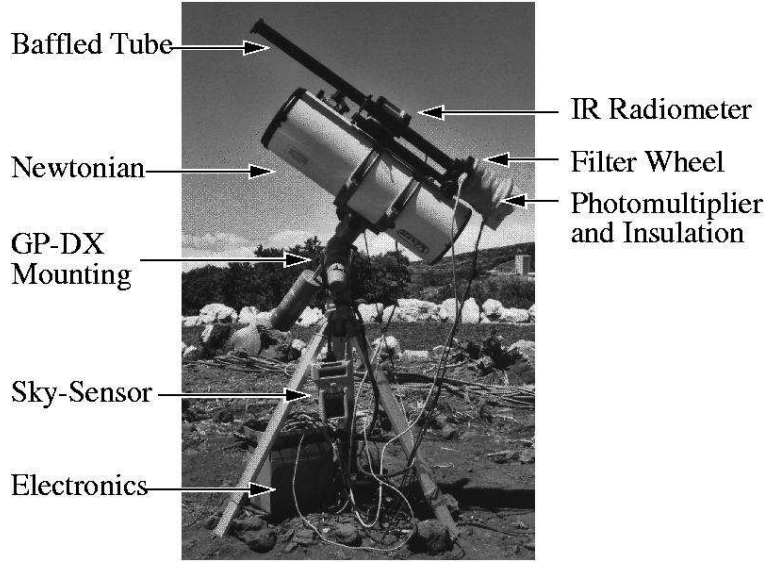}}
\caption{Setup to measure the night sky background light, mounted on a computer-controlled
amateur telescope for pointing control.}
\label{fig_sys}
\end{center}
\end{figure}

\begin{figure}
\begin{center}
\mbox{
\epsfxsize7.0cm
\epsffile{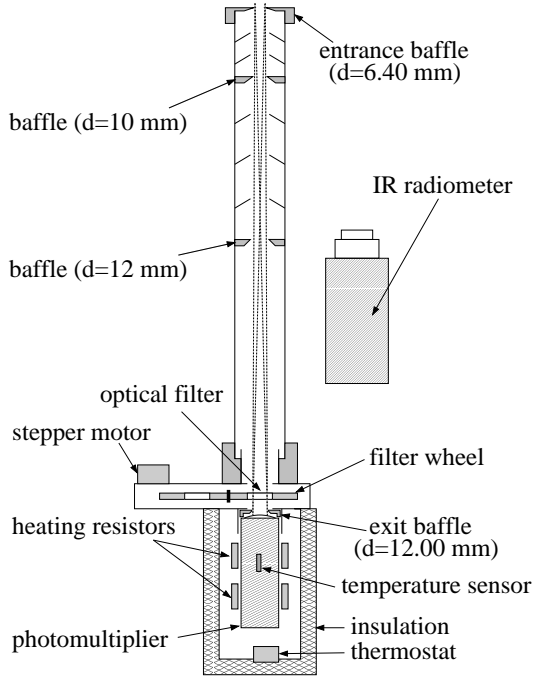}}
\caption{The instrument consisting of a baffled
tube defining the solid angle, a remote-controlled filter wheel and a 
temperature-controlled PMT module including a discriminator and counter.}
\label{fig_instr}
\end{center}
\end{figure}

Fig.~\ref{fig_sys} and Fig.~\ref{fig_instr} illustrate the design of the
instrument. Basically, it consists of a baffled tube which 
defines the solid angle viewed by a
photomultiplier. A remote-controlled filter wheel in front of
the PMT allows to swap in the four narrow-band filters, and to cover the
PMT to determine the rate of dark counts. The whole setup is 
mounted on a computer-controlled amateur telescope, to control
the pointing
(a VIXEN
\footnote{VIXEN Optical Industries LTD.,
247 Hongo, Tokorozawa, Saitama 359-0022, Japan, www.vixen.co.jp} Newtonian with 800~mm focal length and f/d = 4 on
a GP-DX mount interfaced via a Skysensor 2000 PC to a portable
computer running Linux). 
The setup furthermore includes an IR radiometer, for the 
purpose to detect clouds resulting in a increased IR sky
temperature \cite{BUC98}. 

As PMT, the Hamamatsu HC 135 module is used, which
includes a PMT with a 21~mm bi-alkali photocathode,
the high voltage supply, an amplifier, discriminator, 
counter, timer and a control microprocessor. The unit is
interfaced via a RS232 port to the portable computer.
The discriminator threshold and HV are factory-preset to
count signals above the ``valley'' of the single-photon
response. The built-in processor automatically corrects 
dead-time during high-rate measurements.

\begin{figure}
\begin{center}
\mbox{
\epsfxsize8.0cm
\epsffile{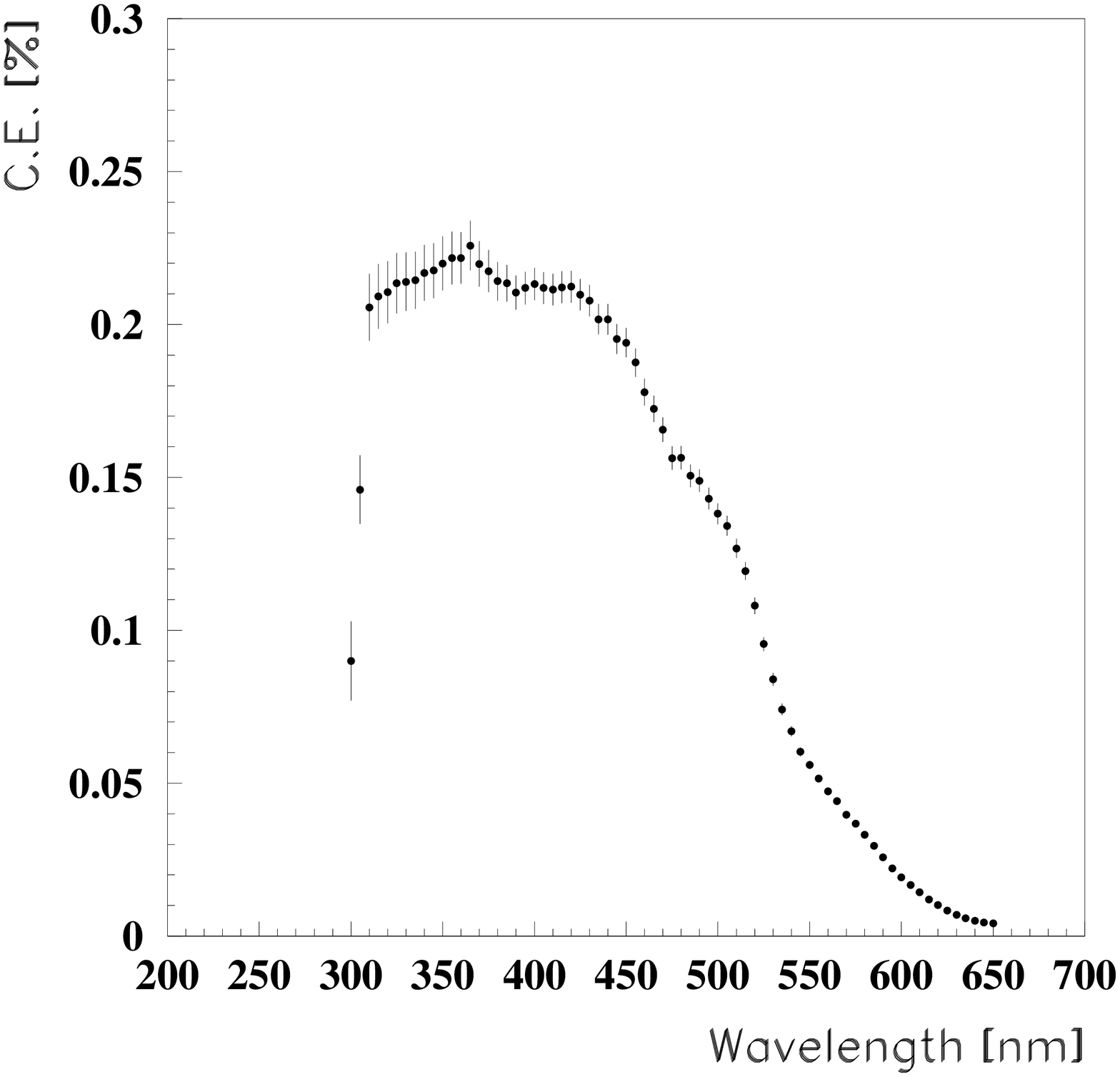}}
\caption{Measured photon counting efficiency of the PMT as a function of wavelength,
using a calibrated silicon photo diode as a reference.
Error bars include systematic errors.}
\label{fig_qe}
\end{center}
\end{figure}

The unit was tested and calibrated in the laboratory.
The setting of the threshold was verified. Illuminating
the device through a calibrated optical attenuator, 
the linear relation between photon flux and
corrected count rate was verified at the 1\% level, for count
rates up to 25 MHz; in the field, measurements
rates varied between 50 Hz and 5000 Hz, depending on the region
of the sky viewed and the filter settings. The photon counting
efficiency of the unit was determined using monochromatic
light, with calibrated Hamamatsu and Newport silicon photo 
diodes serving as a reference, see Fig.~\ref{fig_qe}. In these measurements,
the photocathode was illuminated over a 10~mm diameter, similar 
to the diameter of the final baffle in the collimating tube.
The resulting counting efficiencies are in good agreement
with the calibration data at
450, 550 and 650 nm, provided with the Hamamatsu HC 135. 
The laboratory measurements showed a temperature dependence
of the detected rate between $0.1\%$ and $0.4\%$ per degree Centigrade,
consistent with specifications. In order to eliminate influence
of temperature in the field measurements, the PMT unit was
surrounded by thermal insulation, and thermostat-controlled 
heating resistors were used to stabilize the temperature at
$22 \pm 3$ $^\circ C$. The PMT temperature was monitored during
all measurements.

\begin{figure}
\begin{center}
\mbox{
\epsfxsize8.0cm
\epsffile{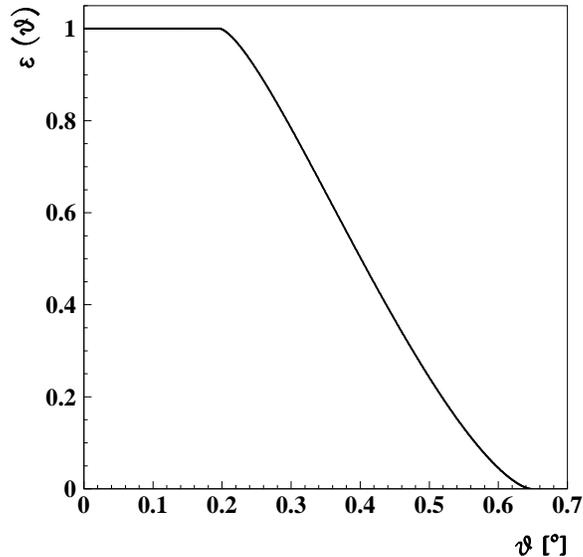}}
\caption{Calculated
transmission of the baffled tube as a function of the angle relative to the 
axis, normalized to the amount of light incident through the entrance
aperture.}
 \label{fig_eff}
\end{center}
\end{figure}

The baffled tube defining the solid angle viewed by the PMT has an
entrance baffle of 6.40~mm diameter, an exit baffle of 12.00~mm,  and
a number of intermediate baffles to reduce the influence of stray light.
Fig.~\ref{fig_eff} shows the calculated transmission
$\epsilon(\vartheta)$ as a function of the
angle $\vartheta$ to the axis. For the calculation a point light source in the
infinite is assumed and $\epsilon (\vartheta)$ is normalized to the amount
of light incident through the entrance aperture.
Integrating the transmission curve over the solid angle, 
$$
\Omega = 2 \pi \int \epsilon(\vartheta) \sin(\vartheta) \mbox{d}\vartheta 
$$
one finds an effective solid angle coverage $\Omega$ of 
$(1.70 \pm 0.02) \cdot 10^{-4}$ sr, using the geometrical dimensions given
above and their errors. The effective solid angle was measured by first
illuminating the bare PMT using an Ulbricht sphere 
with apertures of defined size in front of the PMT and the Ulbricht sphere
to provide a light
source of well-defined solid angle; the counting rate was determined
as a function of solid angle. 
By measuring the rate after adding the baffle system with the Ulbricht sphere 
positioned to illuminate the entire field of view, the effective solid
angle was then determined.
From these measurements, an effective solid angle
of $\Omega_{eff}=(1.79 \pm 0.05) \cdot 10^{-4}$ sr was derived, in reasonable agreement
with the geometrical estimate given above. In the following, the measured
value is used; it is equivalent to an effective diameter of
the field of view of $0.87^\circ$. 
The influence of stray light was studied by illuminating
the baffled tube from different angles, and by pointing the device at a 
large illuminated area, where only the central spot -- corresponding to
the nominal aperture of the baffled tube -- was blackened. After a number
of iterations concerning the placement of baffles, the transmission of stray
light at angles of a few degrees from the optical axis
was reduced to about $10^{-5}$ of the on-axis response,
corresponding to a contribution of around 1\% of the signal
after integration over solid angle, which should be completely
uncritical. The baffled tube was aligned parallel to the telescope axis
to better than $0.5^\circ$.

The six-slot filter wheel (4 color filters, one open slot, one closed
slot) is mounted at the end of the baffled tube, just before the final
baffle in front of the PMT. The transmission as a function
of wavelength was measured; the center wavelengths of the filters were
determined to 360, 399, 448, and 498~nm. The transmission
is well described by Gaussians of width 4.3, 4.1, 3.5, and 4.3 nm, respectively.
To relate count rate and photon flux, the measured transmission curves
were used.

In order to test the long-term stability of the instrument and its
calibration, a stable laboratory light source was employed. Measurements
proved reproducible within $0.5\%$, even after the instrument was 
completely disassembled and reassembled several times.

To monitor and detect cloud layers during the measurements,
an IR radiometer (Heitronics KT15.82D) was integrated into the setup.
The instrument senses IR radiation in the $8~\mu$m to $14~\mu$m
window and converts the yield into an effective sky temperature, 
within the limits of $-60^\circ$C to $500^\circ$C. The field of view
of about $2^\circ$ is defined by the size of the sensor combined with
a germanium lens. The use of IR radiometers in sky monitoring for
Cherenkov telescopes has been introduced and discussed by the
Durham group \cite{BUC98}.

In addition, the 
ambient temperature and the humidity were monitored during
all measurements.

\section{Determination of photon flux}
\label{secflux}

A measurement results in a photon counting rate, usually
averaged over 10~s. At count rates between 50~Hz and 5000~Hz,
depending on the filter settings, statistical errors are
small. Dark counts -- typically at a level of 20~Hz -- are
subtracted, but usually represent only a minor correction.

For the measurements with a narrow-band filter
with center wavelength $\lambda$,
the measured rate $R$ can be directly converted into a photon
flux 
$$
\phi(\lambda) = {\mbox{d}N \over \mbox{d}A \mbox{d}\Omega \mbox{d}t \mbox{d}\lambda}
= {R \over A_{in} \Omega_{eff} \int T(\lambda') \epsilon_{PMT}(\lambda')
\mbox{d}\lambda'}
$$
using the measured counting efficiency $\epsilon_{PMT}(\lambda)$, the 
filter transmission $T(\lambda)$, 
and the entrance area $A_{in}$ and effective solid angle $\Omega_{eff}$ of the baffled tube,
under the assumption that the spectral distribution is
approximately constant over the width of the filter response.
Including the uncertainties in the measurements of quantum efficiencies
and filter transmissions, the typical systematic errors of
$\phi(\lambda)$ vary between 6.4\% at 360~nm and 4.5\% at 500~nm.

\begin{figure}
\begin{center}
\mbox{
\epsfxsize8.0cm
\epsffile{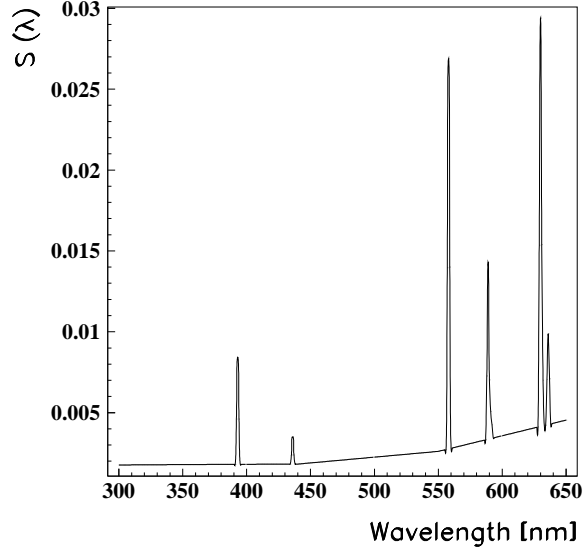}}
\caption{Spectrum assumed in the calculation of the spectrum-averaged
quantum efficiency of the PMT, following \cite{BEN98}.}
\label{fig_spectrum}
\end{center}
\end{figure}

The measurements without filter can be presented in two
ways:
\begin{itemize}
\item In terms of the photoelectron rate $\phi_{pe}$ per solid angle and area
for bi-alkali photo cathodes,
$$
\phi_{pe} = 
{\mbox{d}N_{pe} \over \mbox{d}A \mbox{d}\Omega \mbox{d}t} = 
{R \over A_{in} \Omega_{eff}}~~~. 
$$
This rate can be directly used to
calculate the rate of night-sky photoelectrons in the PMTs of
Cherenkov telescopes, after small corrections to account for possible differences
in the peak quantum efficiency, the collection efficiency and
the fraction of photoelectrons cut below the trigger threshold of 
the HC 135 unit. 
\item In terms of the integral photon rate over a certain wavelength
interval, typically 300~nm to 650~nm, 
$$
\phi = {\mbox{d}N \over \mbox{d}A \mbox{d}\Omega \mbox{d}t}
= {R \over A_{in} \Omega_{eff} <\epsilon_{PMT}>}~~~.
$$
Here, $<\epsilon_{PMT}>$ is the counting efficiency
of the PMT, averaged over the shape of the spectrum $S(\lambda)$
$$
<\epsilon_{PMT}> = \int_{\lambda_1}^{\lambda_2} S(\lambda)
\epsilon_{PMT}(\lambda) \mbox{d} \lambda
$$
where
$$
\int_{\lambda_1}^{\lambda_2} S(\lambda) \mbox{d} \lambda = 1~~~.
$$
The limits $\lambda_1$ and $\lambda_2$ have to be chosen such that
$S(\lambda) \epsilon_{PMT}(\lambda)$ is negligible outside this range.

\end{itemize}
In the second case, the result obviously depends on the 
spectrum $S(\lambda)$ assumed for the night sky background when calculating
$<\epsilon_{PMT}>$. To determine $<\epsilon_{PMT}>$, the spectrum
given in \cite{BEN98} was used. Apart from a number of discrete lines,
this spectrum is approximately flat between 300~nm to 450~nm, and rises
towards longer wavelengths, see Fig.~\ref{fig_spectrum}. The resulting
average quantum efficiency is 9.8\%.
Including a 10\% uncertainty in the shape of the assumed (normalized)
spectrum, we find a 
systematic error of 8.8\% for the integral flux. The rather extreme assumption
of a constant spectrum $S(\lambda) = const$ would result in a 30\% difference
in the derived integral photon flux.

Alternative units used in the astronomical community to describe the 
level of the night sky background light are S10, the equivalent number of 10th
magnitude stars per square degree, or the unit (magnitudes/arcsec$^2$).
The following relations apply for the conversion \cite{Leinert}:
$$
\phi [1/\mbox{sr s m$^2$ nm}] = 4.80 \cdot 10^7 \phi [\mbox{S10 units}]
$$
in the B band (440 nm), and 
$$
\phi [1/\mbox{sr s m$^2$ nm}] = 3.26 \cdot 10^7 \phi [\mbox{S10 units}]
$$
in the V band (550 nm). One S10 unit in turn corresponds to
27.28 mag/arcsec$^2$.

\section{Measurements of the night-sky photon flux at La Palma and Namibia}

The field measurements were carried out in two campaigns at La Palma and 
in the Khomas Highland of Namibia. The La Palma site of the HEGRA Observatory,
on the site of the Observatorio del Roque de los Muchachos, is located 2200 m asl
at $17^\circ 53' 24''$ W, $28^\circ45'34''$ N. The measurements were carried 
out between May 18 and May 30, 2000 under good weather conditions, at 
temperatures between $5^\circ$ and $15^\circ$, and humidities between 20\% and
40\%.
The measurements in Namibia were carried
out between June 23, 2000 and July 3, 2000
on the site of the farm Hakos, about 14 km from the future site of the H.E.S.S.
installation, at 1800 m asl, $16^\circ 21' 28''$ E, $23^\circ 14' 08''$ S.
The sky was clear, and temperatures and humidity were in the same range as for
the La Palma measurements.

\begin{figure}
\begin{center}
\mbox{
\epsfxsize8.0cm
\epsffile{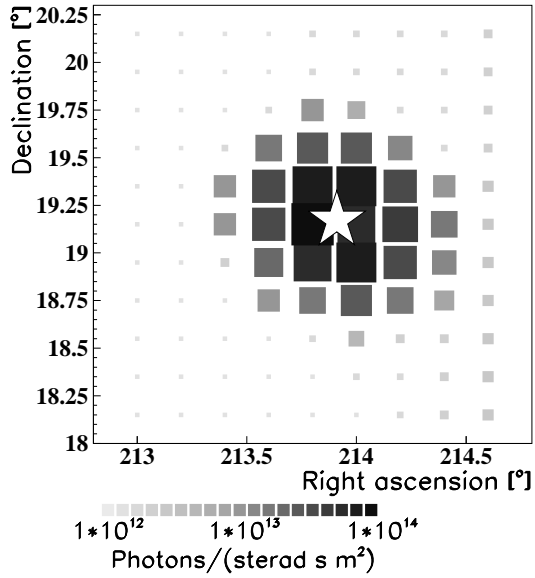}}
\caption{Measured photon flux in the proximity of the 
star Arcturus. The star in the figure
denotes the nominal position of Arcturus.}
\label{fig_star}
\end{center}
\end{figure}

Before each sequence of measurements, the telescope was aligned, and the
alignment calibrated using three reference stars. Typical pointing precision
is in the range of a few arcmin. The pointing was checked moving the instrument 
across a bright star, see Fig.~\ref{fig_star}. The results confirmed both
the pointing and the angular response of the instrument.

Unless otherwise mentioned, all integral measurements quoted in the
following refer to the
wavelength range 300~nm to 650~nm. The photoelectron count rate
is obtained by multiplying the flux by 0.098.

\begin{figure}
\begin{center}
\mbox{
\epsfxsize10cm
\epsffile{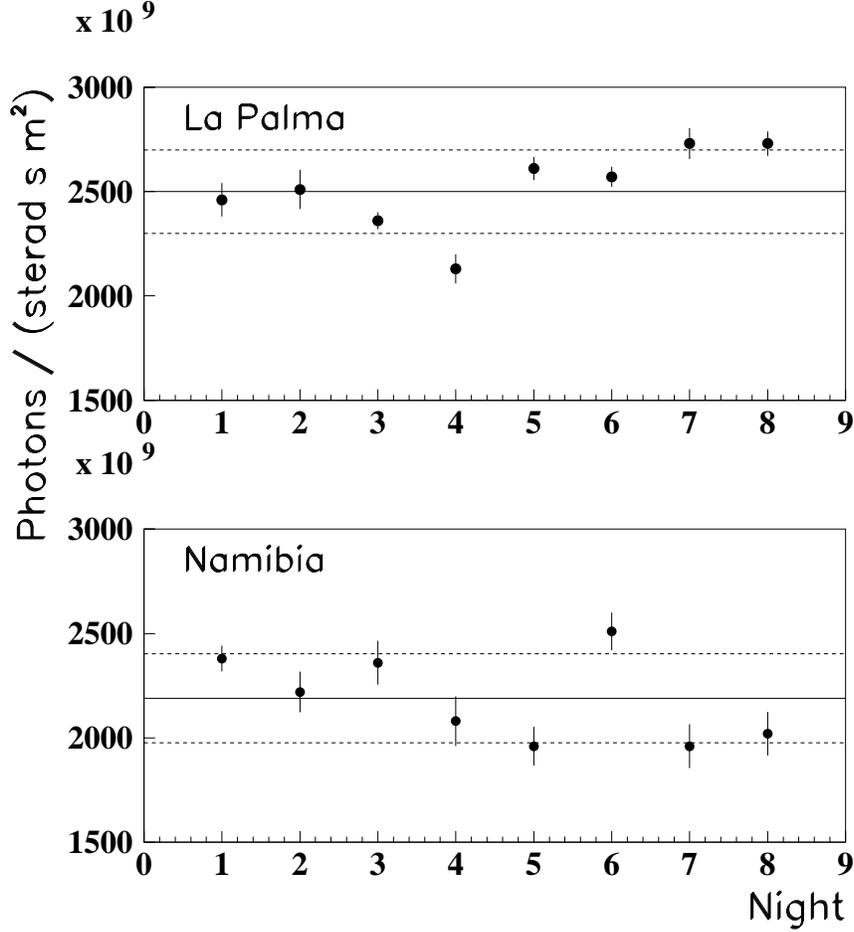}}
\end{center}
\caption{Measurement of the integral photon flux from a set of dark
regions near the zenith, averaged over several
measurements for each single clear night, for La Palma and Namibia.
The full line indicates the mean flux, the dashed lines a band of
$\pm 200 \cdot 10^{9}$ photons/sr s m$^2$.}
\label{fig_rate1}
\end{figure}

\begin{figure}
\begin{center}
\mbox{
\epsfxsize11.0cm
\epsffile{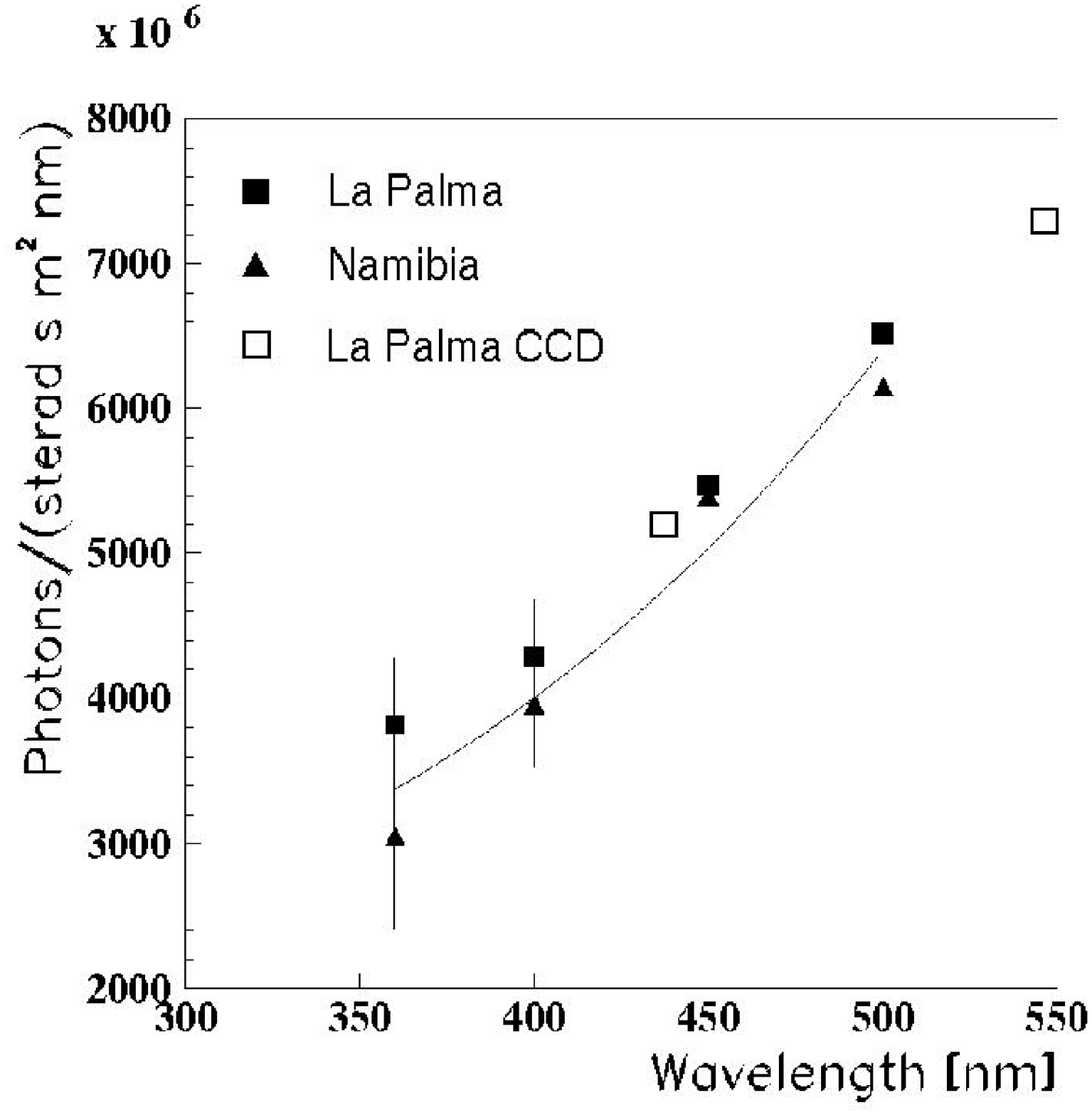}}
\caption{Average 
differential photon flux in
La Palma (full squares) and Namibia (triangles) from a set of dark
regions near the zenith, as a function of wavelength.
Also shown are the values of \cite{BEN98} for La Palma (open
squares), see text for details.}
\label{fig_rate2}
\end{center}
\end{figure}

%{\bf Photon flux from dark regions of the sky}
%
To determine the stability of the measured night sky brightness, a set of dark
regions of the sky near the zenith was measured in several nights, and the
results were averaged for each night. 
Fig.~\ref{fig_rate1}
shows the  photon flux determined for 8 nights of the La Palma measurements, and
for 8 nights of the Namibia campaign. The results show a scatter of about
$\pm 10\%$ between nights. Since the night-to-night variation
is larger than the scatter of the different measurements of a given night,
this points to a systematic variation of the sky brightness.
Measurements of the identical location of the
sky over several hours of the same night showed a 3\% rms variation of 
the night-sky rate. The average flux in Namibia is slightly below
the average La Palma flux. Fig.~\ref{fig_rate2} compares the flux at the
two locations as a function of wavelength. The data can be characterized
by a $\lambda^{3.5}$ dependence.

For comparison, two previous measurements of \cite{MIR94} and
\cite{KAU94} specifically conducted in the framework of Cherenkov
applications are summarized in Table~\ref{table_prev}; 
the La Palma measurement of \cite{MIR94} is in very good 
agreement with the values shown in Fig.~\ref{fig_rate1}; the Jammora value
of \cite{KAU94} is higher, but consistent within errors. 

\begin{table}
\begin{center}
\begin{tabular}{|l|c|c|c|c|c|}
\hline
Location & Ref. & Field of view & Wavelength range & Flux & Scaled flux\\
& & (degr.) & (nm) & $10^{12}$ (sr~s~m$^2$)$^{-1}$ & 
$10^{12}$ (sr~s~m$^2$)$^{-1}$ \\
\hline
La Palma & \cite{MIR94} & 0.43 & 300 - 600 & $1.75 \pm 0.4$ & 2.38 \\
Jammora & \cite{KAU94} & $\approx 0.4$ & 310 - 560 & $1.89 \pm 0.5$ & 3.33 \\
This exp., Fig.~\ref{fig_rate1} & & 0.87 & 300 - 650 & & 2.2 - 2.5 \\
\hline
\end{tabular}
\caption{Other measurements of the night-sky brightness under
conditions relevant for Cherenkov applications. The last column
gives the photon flux scaled to the 300 - 650 nm wavelength range
used in this measurement.}
\label{table_prev}
\end{center}
\end{table}

The long-term median night sky brightness near the zenith at La Palma,
determined from CCD images taken with the Isaac Newton and 
Jacobus Kapteyn telescopes, is given in \cite{BEN98} as
22.7 mag/arcsec$^2$ in the B band, and 21.9 mag/arcsec$^2$ in V.
These values translate into $5.2 \cdot 10^9$ and $7.3 \cdot 10^9$ photons
per sr~s~m$^2$~nm, respectively, in excellent agreement with our data 
(Fig.~\ref{fig_rate2}).

%

\begin{figure}
\begin{center}
\mbox{
\epsfxsize12.0cm
\epsffile{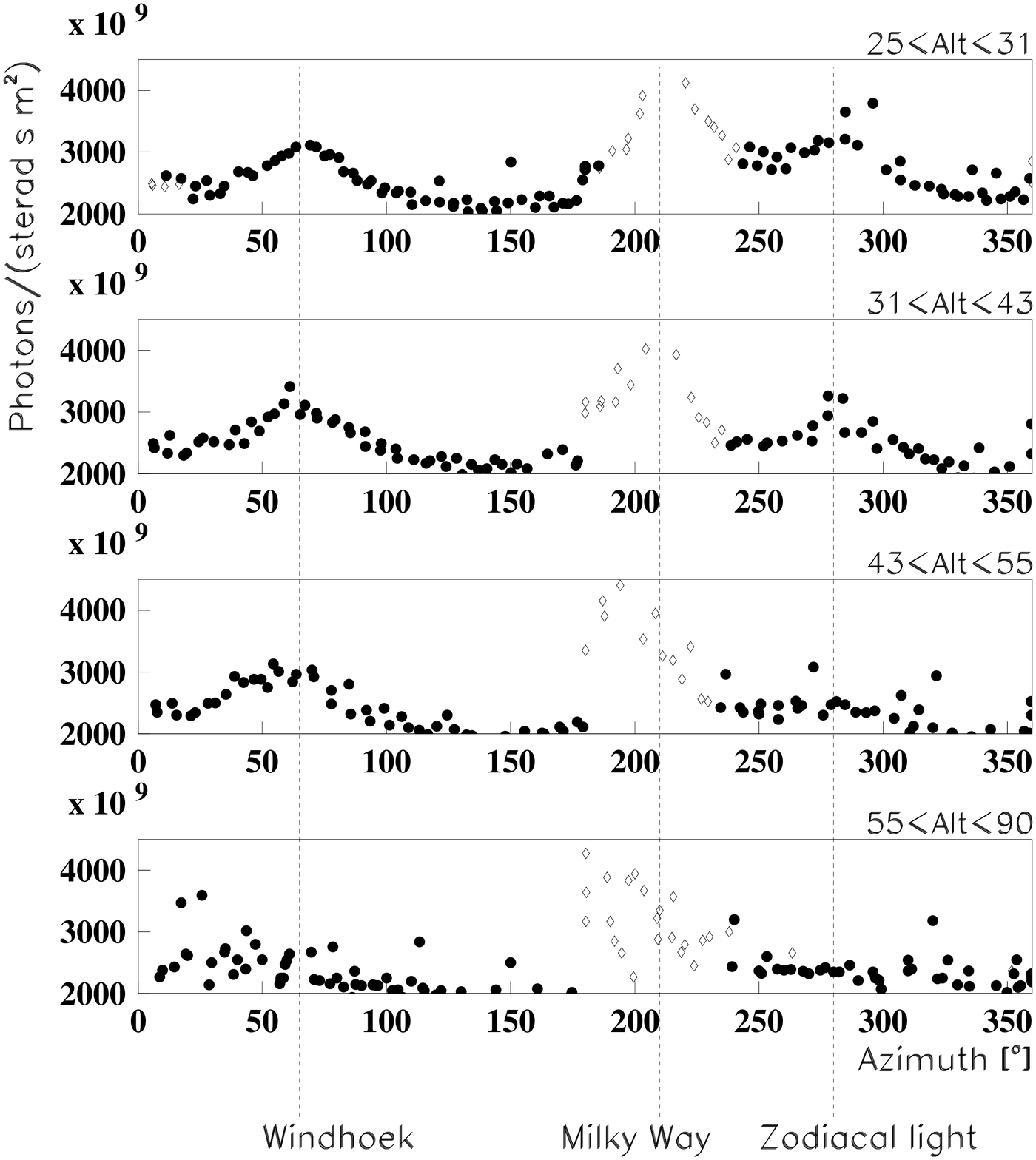}} 
\caption{Photon flux as a function of altitude and azimuth, for Namibia.
Measurements were taken in $5^\circ$ steps in azimuth starting 
at $25^\circ$ altitude; after one
complete turn, the altitude was incremented for the next azimuth scan.
Open symbols denote the region of the Milky Way, with $|b| < 20^\circ$.
Scan points which contain stars brighter than 6 mag are not shown. }
\label{fig_namaltaz}
\end{center}
\end{figure}

%{\bf Dependence ot the flux on altitude and azimuth}
%
The level of the night-sky background light is a function both of 
the terrestrial coordinates azimuth and altitude, and of the celestial
coordinates $b$ and $l$, the galactic latitude and longitude.
Fig.~\ref{fig_namaltaz} illustrates, for the Namibia location,
the variation in the light yield with azimuth and altitude.
In particular at low altitude, one finds a distinct variation
with azimuth. The structures correspond to stray light from the 
city of Windhoek, at a distance of slightly over 100 km, and to
a contribution from zodiacal light. In addition, data points within
the galactic plane (open symbols) show an enhanced flux. The flux
varies by a factor 1.5 outside the galactic plane, and is increased
significantly within the plane.

\begin{figure}
\begin{center}
\mbox{
\epsfxsize13.0cm
\epsffile{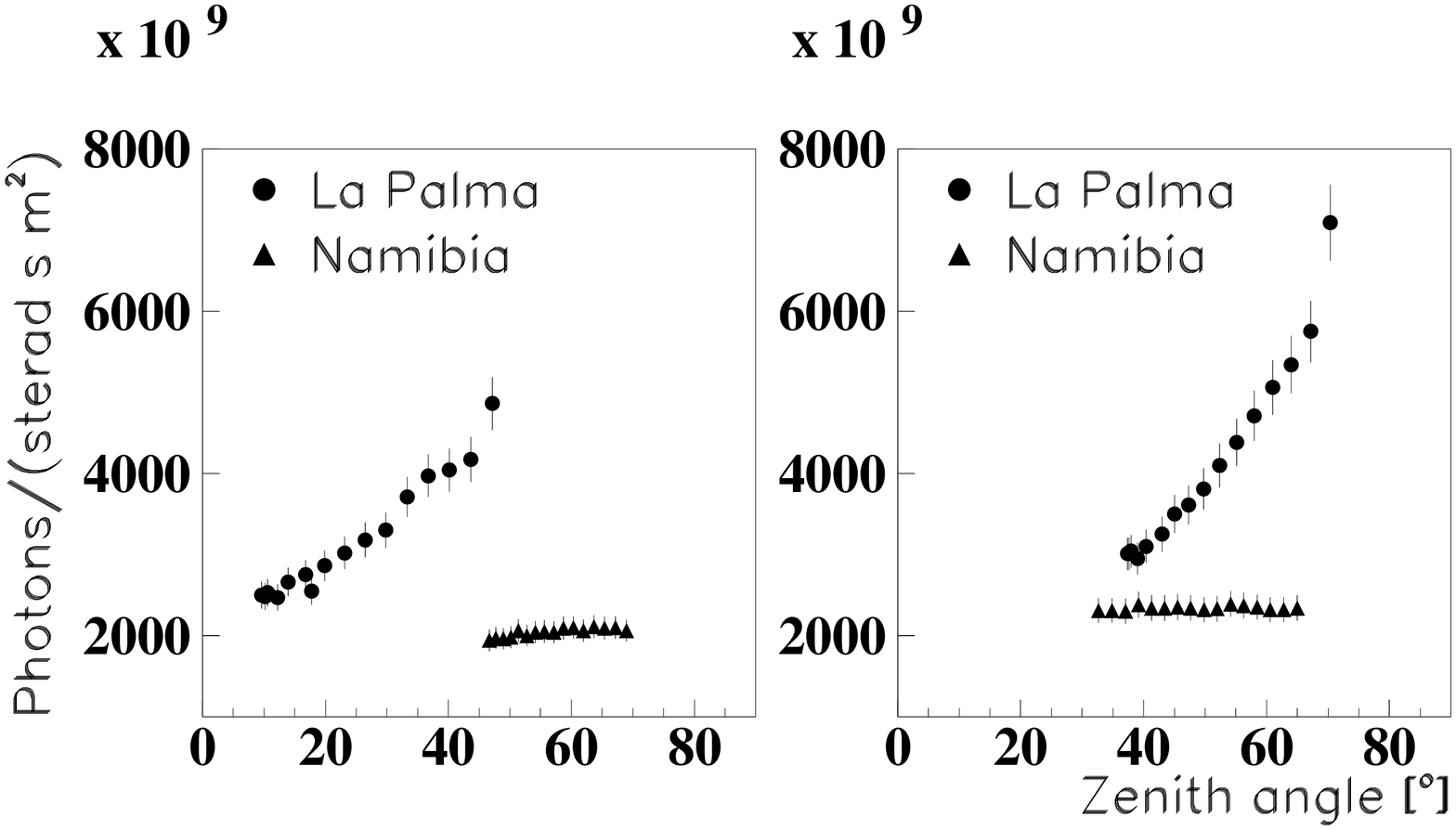}}
% 5.10
\caption{Photon flux from two fixed locations of the sky (l=11.6$^\circ$,
b-71.4$^\circ$, left and l=317$^\circ$, b=53.4$^\circ$, right)
as a function of the zenith angle, for Namibia and La Palma.}
\label{fig_alt}
\end{center}
\end{figure}

Of special interest was the variation of night sky brightness with
altitude; the visual impression was that the sky in Namibia was very
dark and clear down to the lowest altitudes. To separate in one specific
case the dependence on terrestrial and celestial coordinates, two dark
regions of the sky with fixed $b, l$ were measured both from Namibia and
from La Palma, and were followed over a large range in altitude. The
results are summarized in Fig.~\ref{fig_alt}. Whereas at La Palma the 
night sky flux increases with decreasing altitude, the Namibia
data show a constant flux down to relatively low altitude 
\footnote{The altitude-dependence in the La Palma data 
for the two selected regions is somewhat steeper than observed on
average in $alt-az$ scans at La Palma; the scans, however, show 
qualitatively the same feature.}.

%

\begin{figure}
\begin{center}
\epsfxsize8.0cm
\epsffile{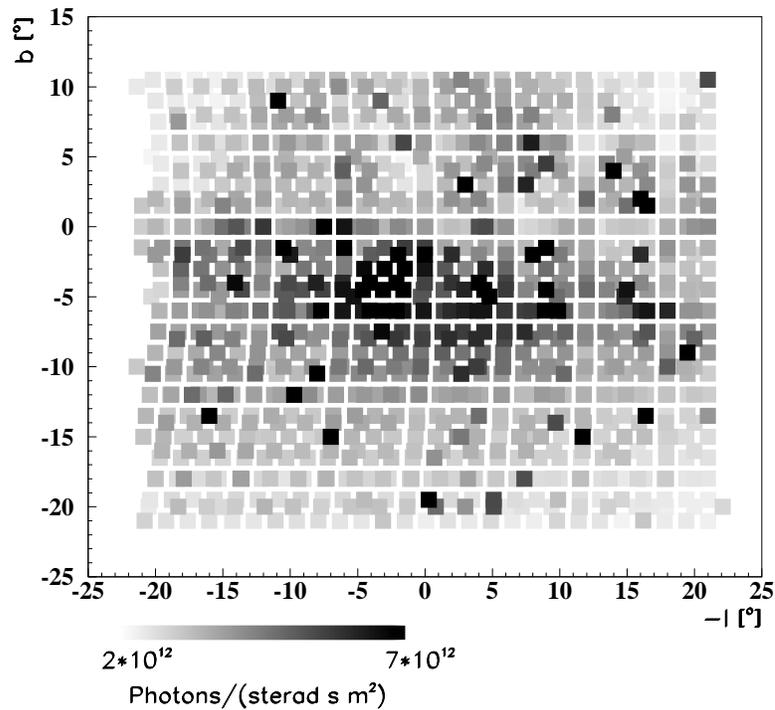}
\end{center}
\caption{Photon flux from the central region of the Galaxy, as a function
of galactic longitude $l$ and latitude $b$. The squares represent 
individual scan points; the grey scale represents photon flux. 
Several scans with different step sizes are combined
in the picture. Isolated points with very high flux indicate bright
stars.}
\label{fig_galplane1}
\end{figure}

\begin{figure}
\begin{center}
\epsfxsize7.0cm
\epsffile{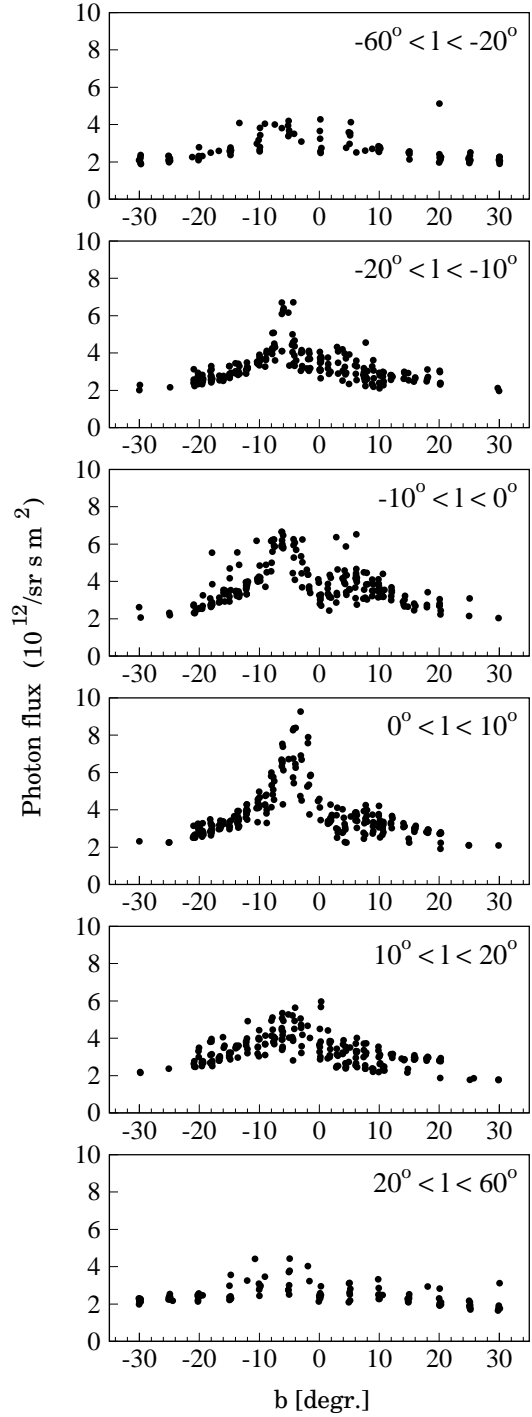}
\end{center}
\caption{Photon flux for different ranges in galactic longitude $l$,
as a function of latitude $b$. Each point represents a scan point;
scans were taken in several nights with a step size of typically
$5^\circ$ at large $b$ and $l$, and $2^\circ$ for the center region.
Point including stars brighter then 6 mag. are not shown.
Statistical errors of the scan points are small; there is a common
8.8\% systematic scale error.}
\label{fig_galplane2}
\end{figure}

%{\bf Photon flux from regions of the Milky Way}
%
In order to explore the variation of night sky brightness across the 
Milky Way, the range $-70^\circ < l < 70^\circ$ and $-30^\circ < b < 30^\circ$
was scanned in several nights, with step sizes varying between
$5^\circ$ for the regions of large $b$ and $l$, to $2^\circ$ for
the region of the galactic center.
Fig. \ref{fig_galplane1}
illustrates the flux variation across the central region of the 
Milky Way.
Fig. \ref{fig_galplane2} shows in more detail the dependence of
the photon flux on $b$, for different ranges of $l$, and also
indicates the scatter of data points in each region.
The galactic center region shows a prominent enhancement of photon
flux, up to a factor 4 above the values obtained for dark regions of
the sky. The flux peaks at $b \approx -5^\circ$; the 
center region at $b = 0, l = 0$ is shadowed by dark clouds. 
Most of the scan points were taken at larger altitudes --
the median altitude is $60^\circ$ -- and the observed variation
of the photon flux should largely reflect a genuine dependence
on $b$ and $l$, as opposed to artefacts resulting from the
variation with $alt$ and $az$ of the zodiacal light and the
light pollution from Windhoek. In particular, data for the center region,
$|l| < 20^\circ$ and small $|b|$, were taken avoiding the 
regions towards Windhoek and towards the zodiacal light. 
Selecting particularly `clean' data sets, using only data
at $alt > 60^\circ$ or taken in the hours around midnight,
the pattern remains unchanged.

Averaging over scan points for altitude $> 60^\circ$ and outside the 
galactic plane, $|b| > 20^\circ$, and removing measurements with 
stars brighter than 6 mag. in the field of view, one finds a mean flux of
$$
\phi = (2.21 \pm 0.22) \cdot 10^{12} ~~\mbox{photons/sr s m$^2$} 
$$
for Namibia, and
$$
\phi = (2.60 \pm 0.35) \cdot 10^{12} ~~\mbox{photons/sr s m$^2$} 
$$
for La Palma,
in the wavelength region from 300~nm to 650~nm. The errors quoted give the
variation between different scan points, based on a Gaussian fit of the
distribution. 
Statistical errors for the individual scan points are small.
As mentioned earlier (Section \ref{secflux}), there is a 8.8\%
systematic error.

\section{Measurement of the atmospheric transmission}

In addition to the night-sky brightness, the atmospheric transmission
was measured by following a star up to large zenith angles. Assuming
constant absorption per unit airmass, the intensity to first order varies with
the zenith angle $\theta_z$ as 
$$
\ln I(\theta_z) = - \tau \sec \theta_z + \ln I_0
$$
where $\tau$ is the optical depth of the (vertical) atmosphere. Within
errors, the transmission was identical on La Palma and in Namibia, 
with typical values of ~60\% at 360~nm, ~72\% at 400~nm, ~80\% at 450 nm
and ~84\% at 500~nm, with errors of 3\% to 4\%.

\section{Photoelectron rate in the PMTs of the 
H.E.S.S. telescopes, including albedo}

The results discussed above show that the observation conditions at the
Namibia site are indeed very good, with the night-sky flux slightly below
the La Palma values for small zenith angles, and a nearly altitude-independent
flux. Observations within the galactic plane and in particular near the
galactic center will have to deal with an increase by a factor 4 in
background levels. For the mirror area of about 94~m$^2$ 
of the H.E.S.S. telescopes (after accounting
for the shadowing of the mirrors due to camera body and camera masts),
and given the typical 80\% mirror reflectivity, the
75\% to 80\% net transmission of the Winston cones in front
of the PMTs, and the $0.16^\circ$ pixel size of the H.E.S.S. cameras,
one derives a background photoelectron rate per pixel
of $85 \pm 13$ MHz for the average
(non-galactic-plane) flux. Additional contributions to the background
rate come from  light rescattered from the ground, and from the structure
of the telescope, which is visible between the round mirror tiles.

Using the identical setup,
and pointing the instrument to the ground,
 the flux of rescattered night-sky background 
light was measured, and is summarized in Table~\ref{tab_scatter}.
Weighting the rescattered night-sky background light from the ground with a reduced 
transmission of the Winston cones at larger angles, one finds that 
ground albedo will contribute a background photoelectron rate of 
$ 13 \pm 4$ MHz. 

Concerning the telescope structure, the flux of backscattered light
depends on the color of the telescope structure. Black color
is obviously prefered, as far as night-sky background is concerned.
On the other hand, white color reduces heating of the telescope
structure during day time; in the sunshine, temperature
differences of $15^\circ$C were measured between metal painted black and white.
For the Cherenkov application, the important quantity is the reflectivity
at short wavelengths, below 500~nm to 600~nm, where the
sensitivity of the PMTs is high. The peak intensity of 
solar radiation heating the telescope structure during day time, on 
the other hand, is at longer wavelengths. A good compromise is therefore
a red color, which absorbs most blue light and which reflects red light.
Table~\ref{tab_scatter} shows that indeed a red color (RAL 3017) is closer
to black than to white as far as the background photoelectron
yield is concerned; on the other hand, the temperature increase 
was measured to only $3^\circ$C, or 20\% of the difference between
black and white.
The actual H.E.S.S. telescopes will be painted in a slightly
darker red, RAL 3016, which should reduce the photoelectron 
yield by another factor 2. Weighted with the relevant solid angle,
light scattered from the telescope structure is expected to contribute
$2.4 \pm 1.2$ MHz of photoelectron rate, resulting in a total estimated
background rate of $100 \pm 13$ MHz per pixel of the H.E.S.S. cameras.

\begin{table}
\begin{center}
\begin{tabular}{|l|c|}
\hline
Material & Diffuse reflected night sky light (Photoel./sr s m$^2$) \\
\hline
Red sand & $(2.2 \pm 0.2) \cdot 10^{10}$ \\
Grass    & $(2.5 \pm 0.2) \cdot 10^{10}$ \\
\hline
Black (RAL 9011) & $(1.3 \pm 0.1) \cdot 10^{10}$\\
White (RAL 9010) & $(13.8 \pm 0.3) \cdot 10^{10}$\\
Rose  (RAL 3017) & $(3.4 \pm 0.1) \cdot 10^{10}$\\
\hline
\end{tabular}
\caption{Diffuse reflected night sky light for different materials.}
\label{tab_scatter}
\end{center}
\end{table}

\section{Summary and conclusions}

Using an instrument developed specially for this purpose, the
level of the night sky background light was surveyed both at
the La Palma site of the HEGRA Cherenkov telescope system, and
near the site of the H.E.S.S. telescope system in Namibia.
Measurements were carried out over typically 10 days at each
location, under good weather conditions. For identical locations
of the sky, measurements show a variation of 10\% between nights.
The average flux of dark regions of the night sky near the zenith 
in La Palma
is $2.6 \cdot 10^{12}$ photons/sr~s~m$^2$, compared to 
$2.2 \cdot 10^{12}$ photons/sr~s~m$^2$ in Namibia, 
for a range between 300 and 650~nm. A remarkable
feature of the Namibia data is that the flux from
dark regions of the sky is independent of the zenith angle,
for zenith angles up to at least $70^\circ$. Since the measurements
essentially represent a snapshot, as compared to an annual average,
it is not a priori clear to which extent long-term or seasonal
variations can account for these differences; on the basis of
visual observations over longer periods, the observation periods
were characterized as representing fairly typical conditions.

Integrated over the field of view of the instrument ($0.87^\circ$),
the night sky background varies significantly with galactic
longitude and latitude. The region near the galactic center shows
fluxes which are increased by a factor 4 compared to dark regions
of the sky, with consequences for the trigger thresholds of
Cherenkov telescopes observing in this region.

The instrument was also employed to study the effect of ground albedo
and reflections from the structure of a Cherenkov telescope; for
Cherenkov cameras where Winston cones properly limit the field of
view of the individual pixels, the albedo contributions to  the
background photoelectron rate of pixel are below 20\%.

\end{document}